\documentclass[aps, pra, longbibliography,twocolumn,showpacs,superscriptaddress]{revtex4-1}  
\usepackage{graphicx}  
\usepackage{dcolumn}   
\usepackage{bm}        
\usepackage{amssymb}   
\usepackage{amsmath}   
\usepackage{hyperref}

\newcommand{\ket}[1]{|{#1}\rangle}
\newcommand{\bra}[1]{\langle{#1}|}

\begin{document}
\title{Diffusion Monte Carlo approach versus adiabatic computation for local Hamiltonians}
\author{Jacob~Bringewatt}
\affiliation{Department of Physics, University of Maryland, College Park, Maryland 20740, USA}
\affiliation{Joint Center for Quantum Information and Computer Science, University of Maryland, College Park, Maryland 20742, USA}
\author{William~Dorland}
\affiliation{Department of Physics, University of Maryland, College Park, Maryland 20740, USA}
\author{Stephen~P.~Jordan}
\affiliation{Joint Center for Quantum Information and Computer Science, University of Maryland, College Park, Maryland 20742, USA}
\affiliation{National Institute of Standards and Technology, Gaithersburg, Maryland 20899, USA}
\author{Alan~Mink}
\affiliation{National Institute of Standards and Technology, Gaithersburg, MD}
\affiliation{Theiss Research, La Jolla, California 92037, USA}    
\date{\today}

\begin{abstract}
Most research regarding quantum adiabatic optimization has focused on stoquastic Hamiltonians, whose ground states can be expressed with only real, nonnegative amplitudes, and thus for whom destructive interference is not manifest. This raises the question of whether classical Monte Carlo algorithms can efficiently simulate quantum adiabatic optimization with stoquastic Hamiltonians. Recent results have given counterexamples in which path integral and diffusion Monte Carlo fail to do so. However, most adiabatic optimization algorithms, such as for solving MAX-$k$-SAT problems, use $k$-local Hamiltonians, whereas our previous counterexample for diffusion Monte Carlo involved $n$-body interactions. Here we present a new 6-local counterexample which demonstrates that even for these local Hamiltonians there are cases where diffusion Monte Carlo cannot efficiently simulate quantum adiabatic optimization. Furthermore, we perform empirical testing of diffusion Monte Carlo on a standard well-studied class of permutation-symmetric tunneling problems and similarly find large advantages for quantum optimization over diffusion Monte Carlo. 
\end{abstract}

\maketitle

Since their introduction~\cite{Farhi} quantum adiabatic algorithms have garnered significant attention for their potential use in solving discrete optimization problems such as the NP-complete maximum satisfiability (MAX-$k$-SAT) problems. For general Hamiltonians quantum adiabatic computation is as powerful as standard quantum computation~\cite{Aharonov}. However, most of the research into this model has dealt with so-called {\it stoquastic} Hamiltonians, for which all off-diagonal matrix elements are real and nonpositive. By the Perron-Frobenius theorem, the ground state of such a Hamiltonian can be expressed by an eigenvector with only real, nonnegative amplitudes \cite{IBM}. As a consequence, the effects of destructive interference are not manifest. This raises the question as to whether quantum adiabatic computation with stoquastic Hamiltonians is capable of exponential speedup over classical algorithms. 

The general complexity-theoretic question of whether such a speed up is possible is still open. However, it is more straightforward to ask whether stoquastic adiabatic computing can outperform specific classical algorithms, such as Monte Carlo. The answer here is yes. Two common Monte Carlo (MC) algorithms are path integral MC and diffusion MC. Hastings demonstrated that due to topological obstructions there is a class of problems where path integral MC fails to efficiently simulate quantum adiabatic computation~\cite{Hastings}. Similarly, Jarret \emph{et al.} presented a problem for which a MAX-$k$-SAT optimized diffusion MC algorithm called Substochastic Monte Carlo (SSMC) fails to efficiently simulate its quantum counterpart~\cite{Jordan}. However, the counterexample of~\cite{Jordan} is highly non-local due to its dependence on a projection operator. Most real applications of quantum adiabatic computation, such as MAX-$k$-SAT, depend on local Hamiltonians. Here we present a 6-local permutation-symmetric Hamiltonian for which SSMC fails to efficiently simulate quantum adiabatic computation. It also appears that a 2-local Hamiltonian presented by Hastings in \cite{Hastings} as a counterexample in which open boundary condition path integral Monte Carlo fails to efficiently simulate stoquastic adiabatic computing may additionally thwart diffusion Monte Carlo for reasons similar to those presented here \footnote{Specifically, the candidate Hamiltonian is from section 2.D of \cite{Hastings} and is made 2-local via the gadgets of section 3 of that paper. We thank Elizabeth Crosson for pointing this out.}.


Substochastic Monte Carlo (SSMC) is a class of diffusion MC algorithms that simulate a time-dependent diffusion process given the same operator as a stoquastic adiabatic process. In the rest of this paper, we will analyze the performance of SSMC compared to stoquastic adiabatic computing. Although our analysis is for SSMC, the results should be generic to most diffusion Monte Carlo algorithms. One exception is diffusion Monte Carlo algorithms that use guiding wavefunctions, which may have drastic effects on performance. However, diffusion Monte Carlo with guiding wavefunctions is a difficult class of algorithms to formalize, because in practice the choice of guiding wavefunction is typically done on an \emph{ad hoc} basis driven by physical intuition.

The SSMC algorithm is described in detail in~\cite{Jordan, jarret2017substochastic} but the core idea is to interpret imaginary-time Schrodinger evolution as a continuous-time random walk for a population of walkers. The generator of this random walk varies with time in accordance with the annealing schedule of the original adiabatic algorithm. This random walk is then discretized into timesteps $\Delta t$ resulting in a Markov chain. Specifically, for sufficiently small $\Delta t$ we can describe the imaginary time evolution of this system by $\psi(t+\Delta t) \simeq \left( I - H(s(t)) \Delta t \right) \psi(t)$. 

For sufficiently small $\Delta t$, the matrix elements $(I-H(s)\Delta t)_{ij}$ are all between 0 and 1 and can be interpreted as the conditional probability for a walker to be at bit string $j$ at the next timestep, given that it is at bit string $i$ at the present timestep. Furthermore, the given probabilities ensure that the quasistationary distribution of the stochastic process is proportional to the quantum ground state. However, in general, $\sum_j (I-H(s)\Delta t)_{ij} < 1$. Thus there is some probability for the walker to ``die''. To prevent exponential decay of the population of walkers, a method of replenishing the population of walkers is necessary. The SSMC algorithm uses an adaptive energy threshold above which walkers are likely to die and below which walkers are likely to spawn new walkers. This threshold is the mean energy of the population with a slight adaptive feedback loop to make sure the population doesn't vary too widely.

A well-converged SSMC simulation simulates the adiabatic process in the sense that the distribution of walkers tracks the probability distribution
\begin{equation}
p_s^{(1)}=\frac{\psi_s(x)}{\sum_{y\in \{0,1\}^n} \psi_s(y)} .
\end{equation}
where $\psi_s$ is the ground state wavefunction of $H(s)$, expressed in the computational basis.


Our 6-local counterexample is of the form

\begin{equation}
H(s)=-\frac{1}{n}\sum_j X_j + V\Big (\sum_j \bar Z_j\Big ) ,
\end{equation}
where $V$ is a sixth-degree polynomial, and $\bar{Z} = (1-Z)/2 = \ket{1}\bra{1}$ so that $\sum_j \bar{Z}_j$ is the operator for Hamming weight, \emph{i.e.} the number of ones in a bit string. Thus, $V\Big (\sum_j \bar Z_j\Big )$ is a 6-local diagonal matrix, which we can think of as a ``potential''.

To analyze permutation-symmetric Hamiltonians like this one we can take advantage of the fact that since $V$ depends only on Hamming weight, $H$ is block diagonal with one $(n+1)\times(n+1)$ dimensional block spanned by the uniform superpositions of bit strings of fixed Hamming weight. These permutation symmetric basis vectors are

\begin{equation}
\ket{\phi_w}=\frac{1}{\sqrt{\binom{n}{w}}}\sum_{|x|=w}\ket{x} .
\end{equation}
where $|x|$ is the Hamming weight of a bitstring $x$. The eigenvalue gap of this $(n+1)\times(n+1)$ Hamiltonian determines the adiabatic runtime since the ground state of $H$ belongs to this subspace and symmetry prevents any transitions out. Therefore we can apply the adiabatic theorem just to this $(n+1)\times(n+1)$ block of the full $2^n\times2^n$ Hamiltonian.

In the $\ket{\phi_w}$ basis the hopping term of the Hamiltonian is a tridiagonal matrix
\begin{eqnarray}\label{eqn:hoppingterm}
-\sum_j X_j \ket {\phi_w} &=&  -\sqrt{(w+1)(n-w)}\ket {\phi_{w+1}} \nonumber \\
&{}&-\sqrt{w(n-w+1)}\ket{\phi_{w-1}}+\ket{\phi_w} .
\end{eqnarray}
The tridiagonal matrix defined by Eq. \ref{eqn:hoppingterm} is similar to a discretized second derivative except that the matrix elements vary as a function of $w$, specifically becoming larger near $n/2$. In particular with no external potential the groundstate is the uniform superposition 
$\sum_w \sqrt{\binom{n}{w}} \ket{\phi_w}=\sum_{x\in V} \ket x$.

We can make a one-dimensional (1D) continuum approximation to the Hamiltonian in the large $n$ limit using $w$ as the continuum variable. We start with $-\frac{1}{2} \frac{d^2}{dw^2}$ and add a fictitious potential, $V_{\mathrm{fict}}$, which captures the effect of the off-diagonal matrix elements becoming larger near Hamming weight $n/2$. This fictitious potential can be represented as a power series, which converges to the exact spectrum as higher order terms are included. This construct can be derived rigorously, along with the specific form of $V_{\mathrm{fict}}$, by performing a formalized Villain transformation as demonstrated in the Appendix of Brady and van Dam~\cite{brady2016spectral}. In our specific example, keeping terms up to sixth order ensures convergence of the continuum approximation at large $n$, as derived later in the paper.

Our potential as a function of Hamming weight $w$ and annealing parameter $s$ is
\begin{equation}
V(w, s)=V_q(w)+V_l(w, s)-V_{\mathrm{fict}}(w)
\end{equation}
where
\begin{equation}
V_q(w)=\frac{\omega^2}{2\delta^2n^2}\Big (\frac{w}{n}-\frac{1}{2}-\frac{\delta}{2}\Big )^4-\frac{\omega^2}{4n^2}\Big (\frac{w}{n}-\frac{1}{2} -\frac{\delta}{2}\Big )^2
\end{equation}
\begin{equation}
V_l(w, s)=\tau (-2s+1)\Big (\frac{w}{n}-\frac{1}{2}-\frac{\delta}{2}\Big )
\end{equation}
\begin{eqnarray}
\label{eqn:vfict}
V_{\mathrm{fict}}(w)=-\frac{2}{n}\Big (\frac{w}{n}-\frac{1}{2}\Big )^2+2\Big (\frac{w}{n}-\frac{1}{2}\Big )^2 \nonumber\\
+2\Big (\frac{w}{n}-\frac{1}{2}\Big )^4+4\Big (\frac{w}{n}-\frac{1}{2}\Big )^6 
\end{eqnarray}

The three parts of this potential are as follows. First is $V_q(w)$, a symmetric quartic double well potential with one well located at Hamming weight $w=n/2$ and another at $w=n/2+\delta n$. The depth of the wells is determined by the parameter $\omega$. 

\begin{figure*}[htb]
\includegraphics[angle=0,width=15cm]{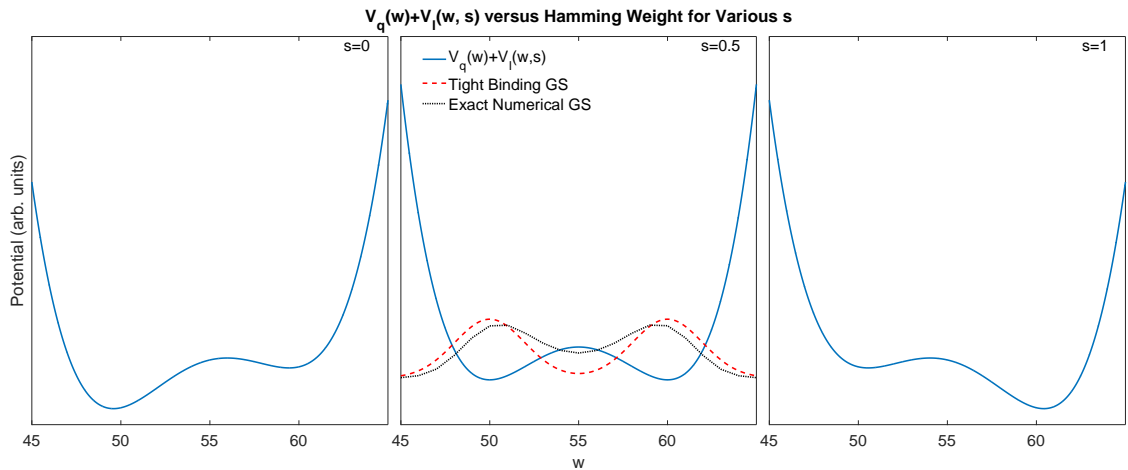}
\caption{Plot of the potential as in the 1D quantum Hamiltonian (with the entropic force canceled by $V_{\mathrm{fict}}$) at various $s$ for $n=100$, $\omega=260\sqrt{n}$, $\delta=0.1(100/n)^{1/4}$, and $\tau=1000/n^{5/4}$. At $s=\frac{1}{2}$ both the exact numerical ground state and the tight binding approximate ground state are shown in the middle panel.}
\label{fig:doublewell}
\end{figure*}

Next, $V_l(w, s)$ is a linear term that varies with $s$ from a positive to a negative slope determined by the parameter $\tau$. It acts to tilt the wells as a function of $s$ so that for $s < \frac{1}{2}$ the left well at $w=n/2$ is energetically favored whereas for $s > \frac{1}{2}$ the right well at $w=n/2+\delta n$ is energetically favored. At $s = \frac{1}{2}$ the linear term has no effect on the quartic double well. The intuitive motivation behind this setup is that when the ground state switches from favoring the left well to favoring the right well SSMC is unable to efficiently track this shift.

The final term, $-V_{\mathrm{fict}}(w)$, is a real potential term judiciously added such that in the continuum approximation it cancels with the resulting $+V_{\mathrm{fict}}(w)$ term. 

Thus, for this potential, we have the following prescription for modeling the qubit Hamiltonian as a 1D continuum problem in the large $n$ limit:
\begin{equation}
-\frac{1}{n}\sum_j X_j + V\Big (\sum_j \bar Z_j\Big ) \rightarrow -\frac{1}{2}\frac{d^2}{dw^2}+V_q(w)+V_l(w, s) .
\end{equation}

In general the parameters $\omega$, $\delta$, and $\tau$ are all functions of $n$. For our specific counterexample we choose $\omega \sim n^{1/2}$, $\delta \sim n^{-1/4}$, and $\tau \sim n^{-5/4}$. We will show that with these scalings the cost for a quantum adiabatic computation with this Hamiltonian will scale polynomially in $n$, whereas the cost for SSMC will scale as $e^{O(\sqrt{n})}$. In addition, we will show that these scalings determine the form of the $V_{\mathrm{fict}}(w)$ term required to ensure convergence.

The performance of quantum adiabatic algorithms can be analytically evaluated by adiabatic theorems~\cite{jansen2007bounds, elgart2012note} which place an upper bound on the runtime of $O(1/\gamma^2)$ for Hamiltonians with a minimum eigenvalue gap $\gamma$ between the ground state and first excited state. We used a tight binding approach in the large $n$ (continuum) limit to determine the scaling of the minimum eigenvalue gap and thus of quantum adiabatic computation for the problem Hamiltonian. 

For this Hamiltonian, the minimum eigenvalue gap will occur at $s = \frac{1}{2}$ when the effective potential is purely the quartic double well ($V_l(w,s=1/2)=0$). For large enough $\omega$ we can treat the ground state wavefunction as the superposition of a left and right wavefunction for each well. In particular, we Taylor series expand to second order the quartic potential that remains following the continuum approximation about the left and right wells at $s = \frac{1}{2}$. The corresponding Gaussian ground states are taken as the left and right wavefunctions $\phi_L$ and $\phi_R$. Then the ground state and first excited state for the double well potential are approximated by $|\phi_{0}\rangle=(1/\sqrt{2})(|\phi_L\rangle+|\phi_R\rangle)$ and $|\phi_{1}\rangle=(1/\sqrt{2})(|\phi_L\rangle-|\phi_R\rangle)$, respectively. The energies of these states are then given by $E_0=\bra{\phi_0}H\ket{\phi_0}$ and  $E_1=\bra{\phi_1}H\ket{\phi_1}$. Then the eigenvalue gap is

\begin{equation}
\gamma\simeq 2\bra{\phi_L}H\ket{\phi_R}\simeq\frac{e^{-\frac{\delta ^2 \omega }{4}} \left(\delta ^2 \omega -3\right)}{8 \delta ^2 n^2}\sim O(n^{-3/2}).
\end{equation}

To see the effectiveness of tight binding consider Fig. \ref{fig:doublewell}. This figure shows the quartic potential as in the 1D quantum Hamiltonian, along with the exact numerical ground state and the tight binding approximate ground state at $s=\frac{1}{2}$. Note that tight binding predicts greater suppresion of the wavefunction in the potential hill between the two wells than there truly is. This is due to the approximation of the quartic wells as a pair of quadratic wells. 

\begin{figure}[htb]
\includegraphics[angle=0,width=9cm]{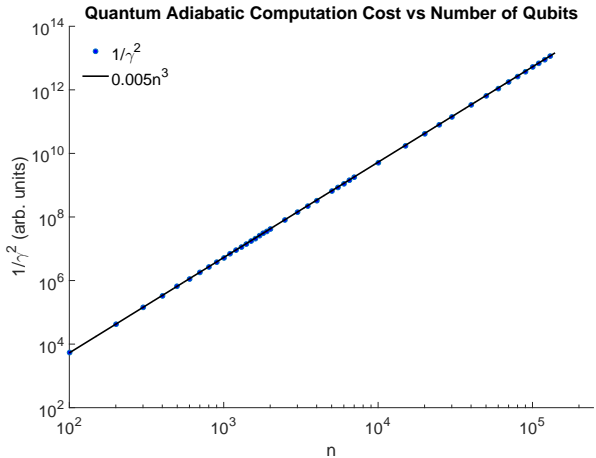}
\caption{A log-log (base 10) scale plot of the cost of quantum adiabatic computation versus the number of qubits with the choice of parameters $\omega=260\sqrt{n}$, $\delta=0.1(100/n)^{1/4}$, and $\tau=1000/n^{5/4}$. Since the ground state of the Hamiltonian lies in the permutation symmetric subspace it can be obtained numerically to a large number of qubits. Like predicted by tight binding the cost scales like $O(n^3)$ for large $n$.}
\label{fig:quantumcost}
\end{figure}

However, the predicted scaling of $\gamma\sim O(n^{-3/2})$ which implies a scaling in cost like $O(1/\gamma^2)\sim O(n^3)$ can be confirmed by direct numerical computation of the minimum eigenvalue gap for the tridiagonal $(n+1)\times(n+1)$ problem Hamiltonian expressed in the permutation symmetric subspace. In this subspace we numerically solved for the eigenvalue gap (and thus the cost) as a function of the number of qubits up to 130,000 qubits. As $n$ becomes large the $s$ value where the minimum gap occurs approaches $s = \frac{1}{2}$ as predicted. Additionally, the cost for quantum adiabatic computation scales like $n^3$ like expected (Fig. \ref{fig:quantumcost}).

With this knowledge of the eigenvalue gap we can derive the form of the $V_{\mathrm{fict}}(w)$ term that arises in the continuum limit (\ref{eqn:vfict}). Consider the kinetic part of the Hamiltonian

\begin{equation}
H_k=-\frac{1}{n}\sum_j X_j=-\frac{\epsilon}{2}\sum_j X_j .
\end{equation}
As shown by Brady and van Dam~\cite{brady2016spectral} this can be rewritten as
\begin{eqnarray}\label{eqn:hk}
H_k=-\frac{1}{2}\bigg [\Big (2\sqrt{1-\hat q^2}+\frac{\epsilon}{\sqrt{1-\hat q^2}}\Big ) -\frac{\epsilon^2 \hat q}{\sqrt{1-\hat q^2}}\hat A \nonumber \\
+\Big (\sqrt{1-\hat q^2}+\frac{\epsilon}{2\sqrt{1-\hat q^2}}\Big )\epsilon^2\hat B\bigg ]
\end{eqnarray}
where $\hat A$ and $\hat B$ are defined so that in the large $n$ (small $\epsilon$) limit they become first and second order derivatives in $q\in [-1, 1]$ respectively, where $q$ is the change of variables
\begin{equation}
q=2\Big (\frac{w}{n}-\frac{1}{2}\Big ) .
\end{equation}
Note that in the continuum limit the operator $\hat q$ becomes the continuous variable $q$.

For our problem the characteristic length scale for the low lying states is the order of $\delta$.  Recall that we chose $\delta\sim O(\epsilon^{1/4})$, $\omega\sim O(\epsilon^{-1/2})$, and  $\tau\sim  O(\epsilon^{3/2})$. So for the low lying states in the wells $O(q\phi)\sim O(\delta)$, and outside the wells the low lying states have $q\phi\approx 0$. We will keep terms in the power series expansion of (\ref{eqn:hk}) up to the same order as the eigenvalue gap, $O(\epsilon^{3/2})$ Note also that with our choice of parameter scaling $V_q\sim V_l \sim O(\epsilon^{3/2})$. In general $\hat A$ and $\hat B$ have operator norms of $O(\epsilon^{-1})$ and $O(\epsilon^{-2})$, respectively. However for low lying states like those we are considering the magnitudes of these operators are determined by the characteristic length scale such that $\hat A\sim O(1/\delta)\sim O(\epsilon^{-1/4})$ and $\hat B\sim O(1/\delta^2)\sim O(\epsilon^{-1/2})$. With this in mind, expanding (\ref{eqn:hk}) to order $\epsilon^{3/2}$ yields

\begin{equation}\label{Hk}
H_k=-1+\frac{\hat q^2}{2}+\frac{\hat q^4}{8}+\frac{\hat q^6}{16}-\frac{\epsilon}{2}\Big (1+\frac{\hat q^2}{2}\Big )-\frac{\epsilon^2}{2}\hat B + O(\epsilon^2).
\end{equation}

The leading order error in this expression is $O(\epsilon^2)$. We can quantify the error induced by neglecting these higher-order terms using perturbation theory. The error in the estimated ground state and ground energy go to zero in the limit that the ratio of the magnitude of the perturbation to the eigenvalue gap of the unperturbed Hamiltonian goes to zero. (See for example \cite{YWS15}.) In our case, the gap is of order $\epsilon^{3/2}$ whereas the leading error term, which we can think of as a perturbation, is of order $\epsilon^2$. So this criterion for convergence to the exact answer is satisfied as $\epsilon \to 0$, \emph{i.e.} $n \to \infty$.

Taking the continuum limit of (\ref{Hk}) and changing variables from $q$ to $w$ yields

\begin{equation}
H_k \rightarrow -\frac{1}{2}\frac{d^2}{dw^2}+V_{\mathrm{fict}}(w)
\end{equation}
where, ignoring constants which don't affect the eigenvalue gap $V_{\mathrm{fict}}$ is given by (\ref{eqn:vfict}) completing the derivation.

While using this framework we showed that quantum adiabatic computation cost scales like $O(n^3)$, SSMC takes superpolynomial time to converge. Although we specifically analyze SSMC, the result should apply more broadly to diffusion Monte Carlo simulations.

\begin{figure*}[!t]
\includegraphics[angle=0,width=15cm]{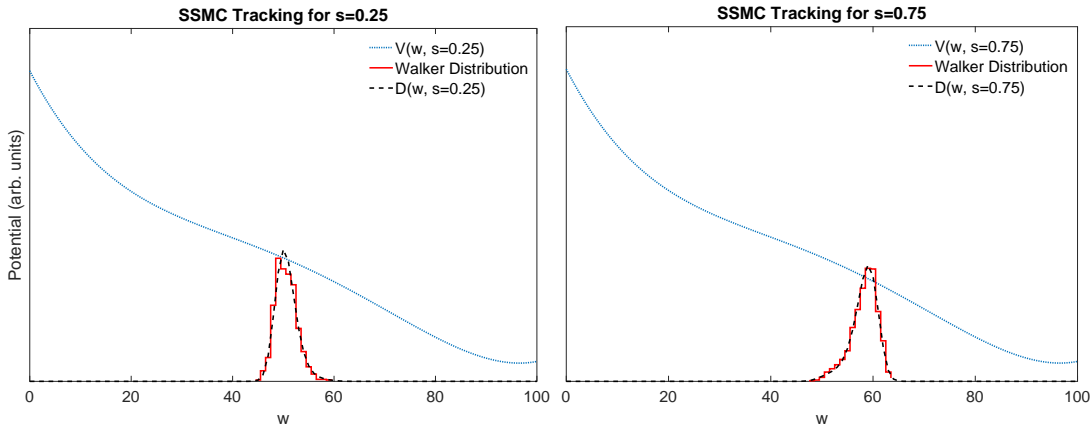}
\caption{Plot of $V(w,s)$ as seen by SSMC for $n=100$ with the choice of parameters $\omega=260\sqrt{n}$, $\delta=0.1(100/n)^{1/4}$, and $\tau=1000/n^{5/4}$. Also shown is the walker distribution for 1000 walkers and 4000 timesteps and the predicted distribution $D(w)$, which is proportional to the quantum ground state in the permutation symmetric subspace. For $s<0.5$ diffusion biased towards $w=n/2$ and teleportation to higher Hamming weight balance to track $D(w)$. For $s>0.5$ SSMC can still track $D(w)$ as depicted, but it requires exponential resources.}
\label{fig:ssmctracking}
\end{figure*}

In SSMC walkers can move either via diffusion or via a sequence of death followed by reproduction at the location of existing walkers with lower potential energy (via the replenishment process previously described). If we assume that for $s< \frac{1}{2}$ SSMC is able to appropriately simulate the ground state wavefunction of the left well $\psi_L$, it is evident that these two processes must oppose one another in such a way that an equilibrium is established. In particular, note that since the walkers see the full potential $V(w,s)$ the fictitious potential dominates at large $n$ such that reproduction will always favor moving towards the location of the right well (see Fig. \ref{fig:ssmctracking}). This is due to the fact that the probability of reproduction is proportional to the difference between a walker's energy and the average energy of the distribution. Diffusion, however, is always biased left towards $w=n/2$ due to the concentration of states around that point. Thus to track the quantum adiabatic computation for $s< \frac{1}{2}$ these two competing effects must be balanced such that the appropriate distribution $p_s^{(1)}$ is maintained. 

However, if SSMC is indeed tracking the quantum ground state for $s< \frac{1}{2}$ it takes a superpolynomial number of walkers for SSMC to track the shift in distribution to the location of the right well for $s> \frac{1}{2}$. In particular, as $s$ increases the potential drops at the location of the right well and reproduction at the location of the right well becomes more likely, but if no walkers are there in the first place reproduction cannot occur. Quantitatively, if the walkers have a distribution proportional to $\psi_L$ the probability of there being walkers in the location of the right well, $P_R$ is given by the proportion of that wavefunction in the right well. That is 

\begin{equation}\label{eqn:ptele}
P_R = \sum_{x \in \{0,1\}^n, |x| \geq n/2+\chi} p^{(1)}(x) = \sum_{w = n/2+\chi}^n D(w)
\end{equation}
where $\chi=\delta n$ is the distance in Hamming weight between the left and right wells and $D(w)$ is the classical probability distribution over Hamming weights
\begin{equation}\label{eqn:D}
D(w) = \frac{ \sum_{|x| = w} \psi_L(x)}{\sum_x \psi_L(x)} =
\frac{\sqrt{\binom{n}{w}} \phi_L(w)}{\sum_{w=0}^n \sqrt{\binom{n}{w}} \phi_L(w)} .
\end{equation}
For large $n$ the binomial factors can be approximated with Gaussians and the sums with integrals. This yields

\begin{equation}
P_R \simeq \frac{1}{2 \sqrt{\pi}} \exp \left[ -
\delta^2 \left( n + \frac{\omega}{2} \right) \right] \sim
\exp [- O(\sqrt{n})].
\end{equation} 

Thus if SSMC tracks $\psi_L$ for $s< \frac{1}{2}$ the probability of there being walkers at the location of the right well so that the algorithm can track the shift in quantum ground state for $s> \frac{1}{2}$ shrinks exponentially with the number of qubits. We see empirically that SSMC does indeed track the expected distribution for $s < \frac{1}{2}$. As a result, SSMC cannot efficiently track for $s > \frac{1}{2}$ and has complexity $e^{O(\sqrt{n})}$.

\begin{figure}[htb]
\includegraphics[angle=0,width=9cm]{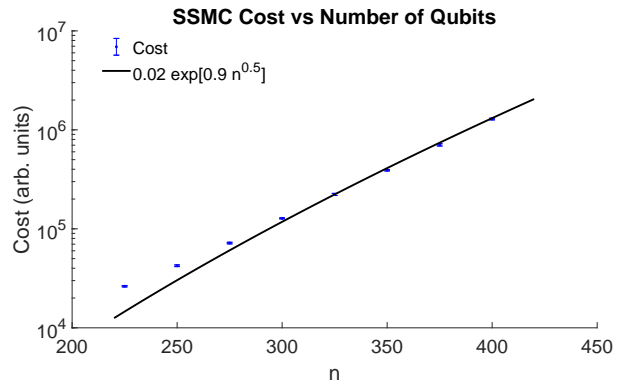}
\caption{Semilogarithmic (base 10) scale plot of the cost of SSMC versus the number of qubits with the same choice of parameters as Fig. \ref{fig:quantumcost}. As predicted this scales superpolynomially with the number of qubits, converging towards the expected $\exp [O(\sqrt{n})]$ scaling as $n$ increases.}
\label{fig:ssmccost}
\end{figure}

We tested SSMC on our problem Hamiltonian with the same choice of parameters $\omega$, $\delta$, and $\tau$ as for the exact numerical determination of the eigenvalue gap. For each number of qubits SSMC was run with a constant number of timesteps (2000) and the number of walkers was varied to obtain a success rate of finding the minimum of the potential to be between 70\% and 75\% percent over 1000 runs. Cost was estimated as being directly proportional to the number of walkers and timesteps, weighted by the probability of success. Cost was calculated ten times for each number of qubits and the mean and standard deviation were used as the value and uncertainty, respectively. Fig. \ref{fig:ssmccost} shows good agreement between observed scaling and the prediction from the tight-binding approximation.

Finally, note the distinct difference in Fig. \ref{fig:ssmctracking} between the full monotonic potential seen by SSMC for which the $-V_{\mathrm{fict}}$ term dominates and the quartic double well potential in the 1D quantum problem for which the $-V_{\mathrm{fict}}$ term is perfectly canceled by the fictitious potential that results from the continuum approximation described above. As a result we emphasize that the key feature of this problem is not that it is hard to solve classically in general. Instead the important point is the inefficiency of classically simulating the quantum adiabatic process using SSMC for this example.

\begin{figure}[htb]
\includegraphics[angle=0,width=9cm]{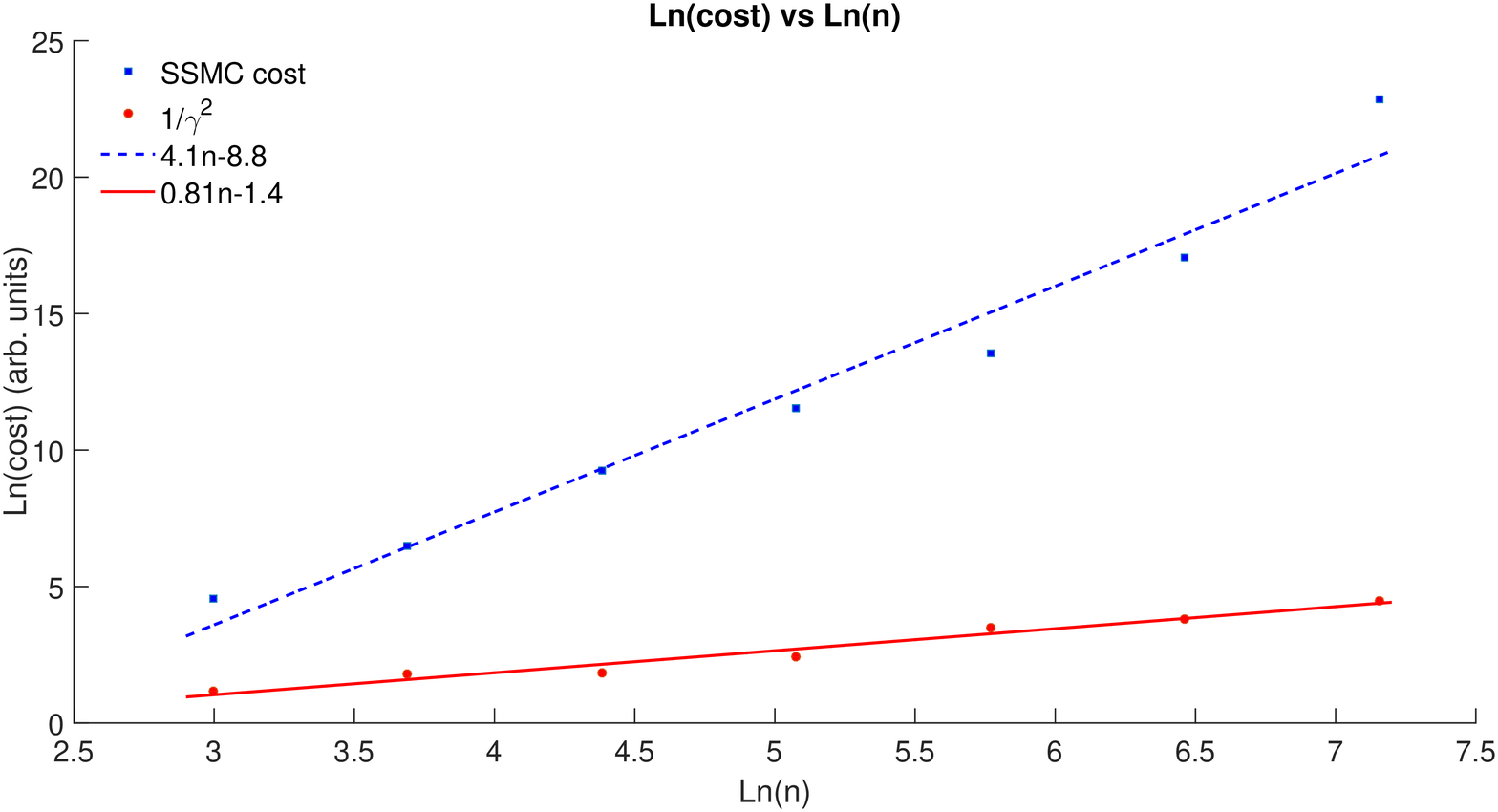}
\caption{Cost of adiabatic optimization compared to the cost of SSMC for the ``spike'' potential from \cite{FGG02, brady2016spectral, R04, CD14, MAL16}. The cost of SSMC is estimated as $(\textrm{No. of walkers}) \times (\textrm{No. of timesteps})/(\textrm{probability of success})$. At the bit numbers we were able to test ($n \leq 1280$), one observes a large discrepancy in scaling between SSMC and adiabatic optimization. Best-fit power-law scalings are shown, illustrating a roughly quintic speedup of adiabatic computing over SSMC. However, both the analysis of \cite{BvD17} and the systematic-looking behavior of the residuals from the fit to the SSMC cost suggest that asymptotic behavior for $n \to \infty$ may differ from the trends seen here. All SSMC trials shown here used 8000 walkers.}
\label{fig:spikeplot}
\end{figure}

The above examples are fine-tuned to illustrate the possibility of superpolynomial discrepancies between the performance of diffusion Monte Carlo and stoquastic adiabatic optimization. It is interesting to also investigate whether such discrepancies arise more generically. To this end, we have compared the performance of adiabatic optimization and SSMC on a standard set of examples that has been previously well studied \cite{FGG02,brady2016spectral, R04, CD14, MAL16}. In these examples, the potential is again a function only of Hamming weight, and consists of a linear potential with a minimum at Hamming weight zero, plus a spike at Hamming weight $n/4$ of height and width $n^\alpha$, which creates a barrier to reaching the global minimum. Prior work \cite{brady2016spectral} has fully characterized the asymptotic behavior of the eigenvalue gap as a function of $\alpha$. However, it was also shown that the asymptotic behavior is not reflected by finite-$n$ examples until extremely large values of $n$, \emph{e.g.} $10^{12}$ \cite{BvD17}. Here we compare the adiabatic performance $1/\gamma^2$ against empirical performance of SSMC at $\alpha = 0.4$ for various $n$. Our results are shown in figure \ref{fig:spikeplot}, and show large discrepancies between the scaling of the quantum algorithm and the Monte Carlo algorithm. 

Our 6-local counterexample and our empirical study of the ``spike'' example both demonstrate cases in which diffusion Monte Carlo is vastly outperformed by stoquastic adiabatic computing. In contrast, SSMC displays good performance on standard benchmarking instances of MAX-$k$-SAT~\cite{Jordan}.  This suggests features like that in our counterexample are not typical of these real world problems. The success of the tight binding approximation used here suggests a path toward obtaining a deeper understanding of problems with many local minima, which may better reflect real world optimization problems such as MAX-$k$-SAT.


\begin{thebibliography}{16}%
\makeatletter
\providecommand \@ifxundefined [1]{%
 \@ifx{#1\undefined}
}%
\providecommand \@ifnum [1]{%
 \ifnum #1\expandafter \@firstoftwo
 \else \expandafter \@secondoftwo
 \fi
}%
\providecommand \@ifx [1]{%
 \ifx #1\expandafter \@firstoftwo
 \else \expandafter \@secondoftwo
 \fi
}%
\providecommand \natexlab [1]{#1}%
\providecommand \enquote  [1]{``#1''}%
\providecommand \bibnamefont  [1]{#1}%
\providecommand \bibfnamefont [1]{#1}%
\providecommand \citenamefont [1]{#1}%
\providecommand \href@noop [0]{\@secondoftwo}%
\providecommand \href [0]{\begingroup \@sanitize@url \@href}%
\providecommand \@href[1]{\@@startlink{#1}\@@href}%
\providecommand \@@href[1]{\endgroup#1\@@endlink}%
\providecommand \@sanitize@url [0]{\catcode `\\12\catcode `\$12\catcode
  `\&12\catcode `\#12\catcode `\^12\catcode `\_12\catcode `\%12\relax}%
\providecommand \@@startlink[1]{}%
\providecommand \@@endlink[0]{}%
\providecommand \url  [0]{\begingroup\@sanitize@url \@url }%
\providecommand \@url [1]{\endgroup\@href {#1}{\urlprefix }}%
\providecommand \urlprefix  [0]{URL }%
\providecommand \Eprint [0]{\href }%
\providecommand \doibase [0]{http://dx.doi.org/}%
\providecommand \selectlanguage [0]{\@gobble}%
\providecommand \bibinfo  [0]{\@secondoftwo}%
\providecommand \bibfield  [0]{\@secondoftwo}%
\providecommand \translation [1]{[#1]}%
\providecommand \BibitemOpen [0]{}%
\providecommand \bibitemStop [0]{}%
\providecommand \bibitemNoStop [0]{.\EOS\space}%
\providecommand \EOS [0]{\spacefactor3000\relax}%
\providecommand \BibitemShut  [1]{\csname bibitem#1\endcsname}%
\let\auto@bib@innerbib\@empty
\bibitem [{\citenamefont {Farhi}\ \emph {et~al.}(2000)\citenamefont {Farhi},
  \citenamefont {Goldstone}, \citenamefont {Gutmann},\ and\ \citenamefont
  {Sipser}}]{Farhi}%
  \BibitemOpen
  \bibfield  {author} {\bibinfo {author} {\bibfnamefont {Edward}\ \bibnamefont
  {Farhi}}, \bibinfo {author} {\bibfnamefont {Jeffrey}\ \bibnamefont
  {Goldstone}}, \bibinfo {author} {\bibfnamefont {Sam}\ \bibnamefont
  {Gutmann}}, \ and\ \bibinfo {author} {\bibfnamefont {Michael}\ \bibnamefont
  {Sipser}},\ }\bibfield  {title} {\enquote {\bibinfo {title} {{Quantum
  computation by adiabatic evolution}},}\ }\href
  {https://arxiv.org/abs/quant-ph/0001106} {\bibfield  {journal} {\bibinfo
  {journal} {arXiv preprint quant-ph/0001106}\ } (\bibinfo {year}
  {2000})}\BibitemShut {NoStop}%
\bibitem [{\citenamefont {Aharonov}\ \emph {et~al.}(2008)\citenamefont
  {Aharonov}, \citenamefont {Dam}, \citenamefont {Kempe}, \citenamefont
  {Landau}, \citenamefont {Lloyd},\ and\ \citenamefont {Regev}}]{Aharonov}%
  \BibitemOpen
  \bibfield  {author} {\bibinfo {author} {\bibfnamefont {Dorit}\ \bibnamefont
  {Aharonov}}, \bibinfo {author} {\bibfnamefont {Wim~van}\ \bibnamefont {Dam}},
  \bibinfo {author} {\bibfnamefont {Julia}\ \bibnamefont {Kempe}}, \bibinfo
  {author} {\bibfnamefont {Zeph}\ \bibnamefont {Landau}}, \bibinfo {author}
  {\bibfnamefont {Seth}\ \bibnamefont {Lloyd}}, \ and\ \bibinfo {author}
  {\bibfnamefont {Oded}\ \bibnamefont {Regev}},\ }\bibfield  {title} {\enquote
  {\bibinfo {title} {{Adiabatic Quantum Computation Is Equivalent to Standard
  Quantum Computation}},}\ }\href@noop {} {\bibfield  {journal} {\bibinfo
  {journal} {SIAM Journal on Computing.}\ }\textbf {\bibinfo {volume} {37}},\
  \bibinfo {pages} {166} (\bibinfo {year} {2008})}\BibitemShut {NoStop}%
\bibitem [{\citenamefont {Bravyi}\ \emph {et~al.}(2008)\citenamefont {Bravyi},
  \citenamefont {DiVincenzo}, \citenamefont {Oliveira},\ and\ \citenamefont
  {Terhal}}]{IBM}%
  \BibitemOpen
  \bibfield  {author} {\bibinfo {author} {\bibfnamefont {Sergey}\ \bibnamefont
  {Bravyi}}, \bibinfo {author} {\bibfnamefont {David~P.}\ \bibnamefont
  {DiVincenzo}}, \bibinfo {author} {\bibfnamefont {Roberto~I.}\ \bibnamefont
  {Oliveira}}, \ and\ \bibinfo {author} {\bibfnamefont {Barbara~M.}\
  \bibnamefont {Terhal}},\ }\bibfield  {title} {\enquote {\bibinfo {title} {The
  complexity of stoquastic local {H}amiltonian problems},}\ }\href@noop {}
  {\bibfield  {journal} {\bibinfo  {journal} {Quantum Information and
  Computation}\ }\textbf {\bibinfo {volume} {8}},\ \bibinfo {pages}
  {0361--0385} (\bibinfo {year} {2008})},\ \bibinfo {note}
  {arXiv:quant-ph/0606140}\BibitemShut {NoStop}%
\bibitem [{\citenamefont {Hastings}(2013)}]{Hastings}%
  \BibitemOpen
  \bibfield  {author} {\bibinfo {author} {\bibfnamefont {M.~B.}\ \bibnamefont
  {Hastings}},\ }\bibfield  {title} {\enquote {\bibinfo {title} {{Obstructions
  to classically simulating the Quantum Adiabatic algorithm}},}\ }\href@noop {}
  {\bibfield  {journal} {\bibinfo  {journal} {Quantum Information and
  Computation}\ }\textbf {\bibinfo {volume} {13}},\ \bibinfo {pages}
  {1038--1076} (\bibinfo {year} {2013})}\BibitemShut {NoStop}%
\bibitem [{\citenamefont {Jarret}\ \emph {et~al.}(2016)\citenamefont {Jarret},
  \citenamefont {Jordan},\ and\ \citenamefont {Lackey}}]{Jordan}%
  \BibitemOpen
  \bibfield  {author} {\bibinfo {author} {\bibfnamefont {Michael}\ \bibnamefont
  {Jarret}}, \bibinfo {author} {\bibfnamefont {Stephen~P}\ \bibnamefont
  {Jordan}}, \ and\ \bibinfo {author} {\bibfnamefont {Brad}\ \bibnamefont
  {Lackey}},\ }\bibfield  {title} {\enquote {\bibinfo {title} {{Adiabatic
  optimization versus diffusion Monte Carlo methods}},}\ }\href@noop {}
  {\bibfield  {journal} {\bibinfo  {journal} {Physical Review A}\ }\textbf
  {\bibinfo {volume} {94}},\ \bibinfo {pages} {042318} (\bibinfo {year}
  {2016})}\BibitemShut {NoStop}%
\bibitem [{Note1()}]{Note1}%
  \BibitemOpen
  \bibinfo {note} {Specifically, the candidate Hamiltonian is from section 2.D
  of \cite {Hastings} and is made 2-local via the gadgets of section 3 of that
  paper. We thank Elizabeth Crosson for pointing this out.}\BibitemShut {Stop}%
\bibitem [{\citenamefont {Jarret}\ and\ \citenamefont
  {Lackey}(2017)}]{jarret2017substochastic}%
  \BibitemOpen
  \bibfield  {author} {\bibinfo {author} {\bibfnamefont {Michael}\ \bibnamefont
  {Jarret}}\ and\ \bibinfo {author} {\bibfnamefont {Brad}\ \bibnamefont
  {Lackey}},\ }\bibfield  {title} {\enquote {\bibinfo {title} {{Substochastic
  Monte Carlo Algorithms}},}\ }\href {https://arxiv.org/abs/1704.09014}
  {\bibfield  {journal} {\bibinfo  {journal} {arXiv preprint arXiv:1704.09014}\
  } (\bibinfo {year} {2017})}\BibitemShut {NoStop}%
\bibitem [{\citenamefont {Brady}\ and\ \citenamefont {van
  Dam}(2016)}]{brady2016spectral}%
  \BibitemOpen
  \bibfield  {author} {\bibinfo {author} {\bibfnamefont {Lucas~T}\ \bibnamefont
  {Brady}}\ and\ \bibinfo {author} {\bibfnamefont {Wim}\ \bibnamefont {van
  Dam}},\ }\bibfield  {title} {\enquote {\bibinfo {title} {{Spectral-gap
  analysis for efficient tunneling in quantum adiabatic optimization}},}\
  }\href@noop {} {\bibfield  {journal} {\bibinfo  {journal} {Physical Review
  A}\ }\textbf {\bibinfo {volume} {94}},\ \bibinfo {pages} {032309} (\bibinfo
  {year} {2016})}\BibitemShut {NoStop}%
\bibitem [{\citenamefont {Jansen}\ \emph {et~al.}(2007)\citenamefont {Jansen},
  \citenamefont {Ruskai},\ and\ \citenamefont {Seiler}}]{jansen2007bounds}%
  \BibitemOpen
  \bibfield  {author} {\bibinfo {author} {\bibfnamefont {Sabine}\ \bibnamefont
  {Jansen}}, \bibinfo {author} {\bibfnamefont {Mary-Beth}\ \bibnamefont
  {Ruskai}}, \ and\ \bibinfo {author} {\bibfnamefont {Ruedi}\ \bibnamefont
  {Seiler}},\ }\bibfield  {title} {\enquote {\bibinfo {title} {{Bounds for the
  adiabatic approximation with applications to quantum computation}},}\
  }\href@noop {} {\bibfield  {journal} {\bibinfo  {journal} {Journal of
  Mathematical Physics}\ }\textbf {\bibinfo {volume} {48}},\ \bibinfo {pages}
  {102111} (\bibinfo {year} {2007})}\BibitemShut {NoStop}%
\bibitem [{\citenamefont {Elgart}\ and\ \citenamefont
  {Hagedorn}(2012)}]{elgart2012note}%
  \BibitemOpen
  \bibfield  {author} {\bibinfo {author} {\bibfnamefont {Alexander}\
  \bibnamefont {Elgart}}\ and\ \bibinfo {author} {\bibfnamefont {George~A}\
  \bibnamefont {Hagedorn}},\ }\bibfield  {title} {\enquote {\bibinfo {title}
  {{A note on the switching adiabatic theorem}},}\ }\href@noop {} {\bibfield
  {journal} {\bibinfo  {journal} {Journal of Mathematical Physics}\ }\textbf
  {\bibinfo {volume} {53}},\ \bibinfo {pages} {102202} (\bibinfo {year}
  {2012})}\BibitemShut {NoStop}%
\bibitem [{\citenamefont {Yu}\ \emph {et~al.}(2015)\citenamefont {Yu},
  \citenamefont {Wang},\ and\ \citenamefont {Samworth}}]{YWS15}%
  \BibitemOpen
  \bibfield  {author} {\bibinfo {author} {\bibfnamefont {Y.}~\bibnamefont
  {Yu}}, \bibinfo {author} {\bibfnamefont {T.}~\bibnamefont {Wang}}, \ and\
  \bibinfo {author} {\bibfnamefont {R.~J.}\ \bibnamefont {Samworth}},\
  }\bibfield  {title} {\enquote {\bibinfo {title} {A useful variant of the
  davis-kahan theorem for statisticians},}\ }\href@noop {} {\bibfield
  {journal} {\bibinfo  {journal} {Biometrika}\ }\textbf {\bibinfo {volume}
  {102}},\ \bibinfo {pages} {315--323} (\bibinfo {year} {2015})},\ \bibinfo
  {note} {arXiv:1405.0680}\BibitemShut {NoStop}%
\bibitem [{\citenamefont {Farhi}\ \emph {et~al.}(2002)\citenamefont {Farhi},
  \citenamefont {Goldstone},\ and\ \citenamefont {Gutmann}}]{FGG02}%
  \BibitemOpen
  \bibfield  {author} {\bibinfo {author} {\bibfnamefont {Edward}\ \bibnamefont
  {Farhi}}, \bibinfo {author} {\bibfnamefont {Jeffrey}\ \bibnamefont
  {Goldstone}}, \ and\ \bibinfo {author} {\bibfnamefont {Sam}\ \bibnamefont
  {Gutmann}},\ }\bibfield  {title} {\enquote {\bibinfo {title} {Quantum
  adiabatic evolution algorithms versus simulated annealing},}\ }\href
  {https://arxiv.org/abs/quant-ph/0201031} {\bibfield  {journal} {\bibinfo
  {journal} {arXiv:quant-ph/0201031}\ } (\bibinfo {year} {2002})}\BibitemShut
  {NoStop}%
\bibitem [{\citenamefont {Reichardt}(2004)}]{R04}%
  \BibitemOpen
  \bibfield  {author} {\bibinfo {author} {\bibfnamefont {Ben~W.}\ \bibnamefont
  {Reichardt}},\ }\bibfield  {title} {\enquote {\bibinfo {title} {The quantum
  adiabatic optimization algorithm and local minima},}\ }in\ \href@noop {}
  {\emph {\bibinfo {booktitle} {Proceedings of the 36th annual ACM Symposium on
  Theory of Computing (STOC)}}}\ (\bibinfo {year} {2004})\ pp.\ \bibinfo
  {pages} {502--510}\BibitemShut {NoStop}%
\bibitem [{\citenamefont {Crosson}\ and\ \citenamefont {Deng}(2014)}]{CD14}%
  \BibitemOpen
  \bibfield  {author} {\bibinfo {author} {\bibfnamefont {Elizabeth}\
  \bibnamefont {Crosson}}\ and\ \bibinfo {author} {\bibfnamefont {Mingkai}\
  \bibnamefont {Deng}},\ }\bibfield  {title} {\enquote {\bibinfo {title}
  {Tunneling through high energy barriers in simulated quantum annealing},}\
  }\href {https://arxiv.org/abs/1410.8484} {\bibfield  {journal} {\bibinfo
  {journal} {arXiv:1410.8484}\ } (\bibinfo {year} {2014})}\BibitemShut
  {NoStop}%
\bibitem [{\citenamefont {Muthukrishnan}\ \emph {et~al.}(2016)\citenamefont
  {Muthukrishnan}, \citenamefont {Albash},\ and\ \citenamefont
  {Lidar}}]{MAL16}%
  \BibitemOpen
  \bibfield  {author} {\bibinfo {author} {\bibfnamefont {Siddharth}\
  \bibnamefont {Muthukrishnan}}, \bibinfo {author} {\bibfnamefont {Tameem}\
  \bibnamefont {Albash}}, \ and\ \bibinfo {author} {\bibfnamefont {Daniel}\
  \bibnamefont {Lidar}},\ }\bibfield  {title} {\enquote {\bibinfo {title}
  {Tunneling and speedup in quantum optimization for permutation-symmetric
  problems},}\ }\href@noop {} {\bibfield  {journal} {\bibinfo  {journal}
  {Physical Review X}\ }\textbf {\bibinfo {volume} {6}},\ \bibinfo {pages}
  {031010} (\bibinfo {year} {2016})}\BibitemShut {NoStop}%
\bibitem [{\citenamefont {Brady}\ and\ \citenamefont {van Dam}(2017)}]{BvD17}%
  \BibitemOpen
  \bibfield  {author} {\bibinfo {author} {\bibfnamefont {Lucas~T.}\
  \bibnamefont {Brady}}\ and\ \bibinfo {author} {\bibfnamefont {Wim}\
  \bibnamefont {van Dam}},\ }\bibfield  {title} {\enquote {\bibinfo {title}
  {Discrepancies between asymptotic and exact spectral-gap analyses of quantum
  adiabatic barrier tunneling},}\ }\href@noop {} {\bibfield  {journal}
  {\bibinfo  {journal} {Physical Review A}\ }\textbf {\bibinfo {volume} {95}},\
  \bibinfo {pages} {052350} (\bibinfo {year} {2017})}\BibitemShut {NoStop}%
\end{thebibliography}
\end{document}